\documentclass[11pt]{article}

\usepackage{a4wide}
\usepackage{palatino}

\usepackage{xypic}
\xyoption{curve}
\usepackage{url,amsmath,amssymb,cite,enumitem}

\input{refocusing.mac}

\title{
  A Deforestation of Reducts: Refocusing
}

\author{
  Olivier Danvy \\[1mm]
  Yale-NUS College \& School of Computing \\
  National University of Singapore \\
  \texttt{danvy@acm.org}
}

\date{February 25, 2023}

\begin{document}

\maketitle

\begin{center}\textbf{Abstract}\end{center}
\begin{quote}
In a small-step semantics with a deterministic reduction strategy,
refocusing is a transformation that connects a reduction-based
normalization function (\ie, a normalization function that enumerates the
successive terms in a reduction sequence -- the successive reducts) and a
reduction-free normalization function (\ie, a normalization function that
does not construct any reduct because all the reducts are deforested).
This transformation was introduced by Nielsen and the author in the early
2000's with an informal correctness proof.
Since then, it has been used in a variety of settings, starting with
Biernacka and the author's syntactic correspondence between calculi and
abstract machines, and several formal proofs of it have been put forward.
This article presents a simple, if overdue, formal proof of refocusing
that uses the Coq Proof Assistant and is aligned with the simplicity of
the original idea.
\end{quote}

\noindent
\textbf{Keywords:}
small-step semantics,
Structural Operational Semantics,
context semantics,
reduction semantics,
deterministic reduction strategy,
reduction-based normalization functions,
reduction-free normalization functions,
continuations,
discontinuities,
type isomorphisms,
delimited continuations,
defunctionalization,
contexts,
the Coq Proof Assistant

\vfill

\begin{center}
\end{center}

\vfill

\clearpage

\section{Background And Introduction}
\label{sec:background-and-introduction}

This article motivates and illustrates refocusing through a simple but
telling language of arithmetic expressions with the possibility of errors
(\sectionref{subsec:the-programming-language-of-discourse}),
using the Coq Proof Assistant~\cite{Bertot-Casteran:04}.
This language is specified with a small-step semantics which gives rise
to a reduction-based normalization function that enumerates the reduction
sequence and makes one wish for its reduction-free counterpart
(\sectionref{subsec:small-step-semantics-of-the-programming-language-of-discourse}).
This wish is fulfilled by refocusing the enumeration
from the site of a contracted redex to the site of the next potential redex
without detouring via an intermediate reduct
(\sectionref{subsec:from-reduction-based-evaluation-to-reduction-free-evaluation}).
The structure of the rest of this article is then outlined
(\sectionref{subsec:roadmap}) and its prerequisites and notations are stated
(\sectionref{subsec:prerequisites-and-notations}).

\subsection{The programming language of discourse}
\label{subsec:the-programming-language-of-discourse}

We consider a simple language of arithmetic expressions over natural
numbers with addition and subtraction:
\inputcoq{OPERATOR}
\inputcoq{TERM}
\inputcoq{VALUE}

\noindent
An operation involves an operator and two operands, which are values:
\inputcoq{POTENTIAL_REDEX}

\noindent
Performing an operation yields a term if the given potential redex is an
actual one; otherwise, it yields an error message:
\inputcoq{CONTRACTUM_OR_ERROR}

\noindent
A potential redex where the minuend (\ie, the left operand of a
subtraction) is smaller than the subtrahend (\ie, the right operand of
this subtraction) is not an actual redex, and so contracting it elicits
an error message:
\inputcoq{CONTRACT}

\noindent
Evaluation is achieved by iterating reduction.
It yields either a value (a natural number) or an error message:
\inputcoq{RESULT}

\subsection{Small-step semantics of the programming language of discourse}
\label{subsec:small-step-semantics-of-the-programming-language-of-discourse}

The reduction strategy is leftmost innermost, \ie, depth first and from left to right.
Here are two examples of a reduction sequence of terms in concrete syntax:

\begin{itemize}[leftmargin=3.5mm]

\item
  $(1 + 10) + (2 + 20)$
  $\longrightarrow$
  $11 + (2 + 20)$
  $\longrightarrow$
  $11 + 22$
  $\longrightarrow$
  $33$

\item
  $(1 - (5 + 5)) - (2 - 20)$
  $\longrightarrow$
  $(1 - 10) - (2 - 20)$
  $\longrightarrow$
  ``numerical underflow: $-9$''

\end{itemize}

\noindent
Since in a reduction sequence, several things happen during each
reduction step, let us make them explicit:
{\let\labelstyle=\textstyle
\spreaddiagramrows{0.6cm}
\spreaddiagramcolumns{0.4cm}
 \newcommand{\atadless}{\hspace{-1mm}}
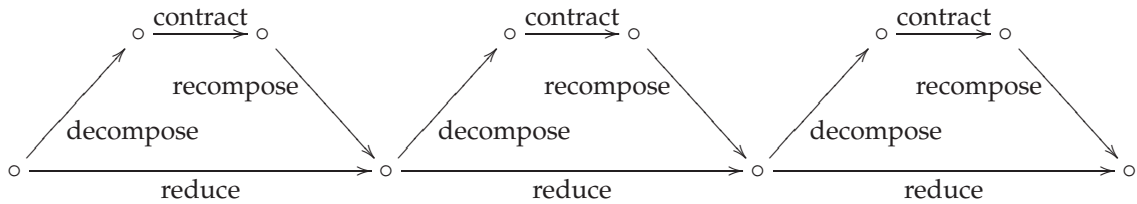
 $$
 \diagram
 &
 \circ
 \rto^{\text{{\small contract}}}
 &
 \circ
 \drto_>>>>>>>>>>>>>>>{\text{{\small recompose}}\atadless}
 &
 &
 \circ
 \rto^{\text{{\small contract}}}
 &
 \circ
 \drto_>>>>>>>>>>>>>>>{\text{{\small recompose}}\atadless}
 &
 &
 \circ
 \rto^{\text{{\small contract}}}
 &
 \circ
 \drto_>>>>>>>>>>>>>>>{\text{{\small recompose}}\atadless}
 \\
 \circ
 \urto_>>>>>>>>>>>>{\atadless\text{{\small decompose}}}
 \ar@{->}[rrr]_{\text{{\small reduce}}}
 &
 &
 &
 \circ
 \urto_>>>>>>>>>>>>{\atadless\text{{\small decompose}}}
 \ar@{->}[rrr]_{\text{{\small reduce}}}
 &
 &
 &
 \circ
 \urto_>>>>>>>>>>>>{\atadless\text{{\small decompose}}}
 \ar@{->}[rrr]_{\text{{\small reduce}}}
 &
 &
 &
 \circ
 \enddiagram
 $$
}

\noindent
In this diagram,
\begin{itemize}[leftmargin=3.5mm]

\item
  the right arrow at the bottom is one-step reduction function that,
  given a reducible term, yields the next reduct in the reduction sequence,

\item
  the slanted up arrow is a decomposition function that, given a
  reducible term, yields a potential redex and its
  context (a context is a term with a hole),

\item
  the horizontal arrow at the top is the contraction function that, given an
  actual redex, yields a contractum (a term), and

\item
  the slanted down arrow is a recomposition function that, given a contractum and a
  context, plugs the contractum in the context and yields the next reduct
  in the reduction sequence.

\end{itemize}

\noindent
Reduction is stuck if the given term cannot be decomposed into a
potential redex and a context or if the potential redex is not an actual
one.
And as captured by the ``there and back'' propositions in
\sectionsref{subsec:SOS-with-two-continuations-and-one-discontinuity:decomposition}{subsec:SOS-with-one-undelimited-continuation-and-one-context-and-one-discontinuity:decomposition},
the recomposition function is a left inverse of the decomposition function.

In the author's mind, this diagram clarifies the
presentation of
reduction in context semantics~\cite[Figure~1]{Felleisen-Hieb:TCS92}:
$$
C[\mathit{r}] \longrightarrow C[c]
$$

\noindent
where $C$ is the context, $\mathit{r}$ is the redex, and $c$ is the contractum,
assuming the contraction rule $r \rightarrow c$.
The turning point for this clarification
was Nielsen candidly pointing out that on the left-hand
side and on the right-hand side of the arrow, the notation $C[\cdot]$
does not mean the same thing.
And indeed on the left-hand side, the notation means that the given term
is decomposed into a reduction context and a potential redex, whereas on
the right-hand side, the notation means that the reduction context is
recomposed around the contractum, yielding the reduct.

The
idea of refocusing is to ``deforest'' -- to quote
Wadler~\cite{Wadler:TCS89} -- the successive reducts in the
reduction sequence by going directly from a contractum and its context to
the next potential redex and its context (or to a value) when the
reduction strategy is deterministic:

{\let\labelstyle=\textstyle
\spreaddiagramrows{0.6cm}
\spreaddiagramcolumns{0.4cm}
 \newcommand{\atadless}{\hspace{-1mm}}
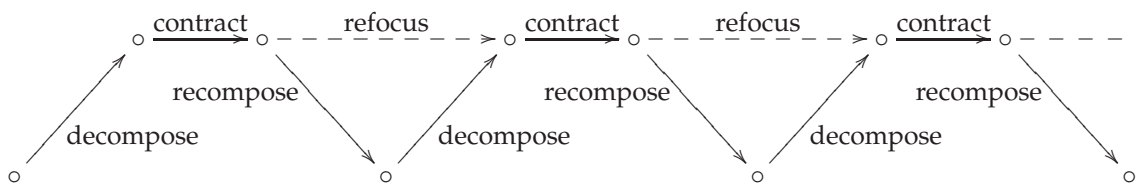
 $$
 \diagram
 &
 \circ
 \rto^{\text{{\small contract}}}
 &
 \circ
 \ar@{-->}[rr]^{\text{{\small refocus}}}
 \drto_>>>>>>>>>>>>>>>{\text{{\small recompose}}\atadless}
 &
 &
 \circ
 \rto^{\text{{\small contract}}}
 &
 \circ
 \ar@{-->}[rr]^{\text{{\small refocus}}}
 \drto_>>>>>>>>>>>>>>>{\text{{\small recompose}}\atadless}
 &
 &
 \circ
 \rto^{\text{{\small contract}}}
 &
 \circ
 \ar@{--}[r]
 \drto_>>>>>>>>>>>>>>>{\text{{\small recompose}}\atadless}
 &
 \\
 \circ
 \urto_>>>>>>>>>>>>{\atadless\text{{\small decompose}}}
 &
 &
 &
 \circ
 \urto_>>>>>>>>>>>>{\atadless\text{{\small decompose}}}
 &
 &
 &
 \circ
 \urto_>>>>>>>>>>>>{\atadless\text{{\small decompose}}}
 &
 &
 &
 \circ
 \enddiagram
 $$
}

\noindent
Evaluating a term by following the reduction sequence (decompose,
contract, recompose, decompose, contract, recompose, etc.)~is referred to
as a ``reduction-based'' evaluation since it enumerates the successive
reducts, and evaluating a term by short-circuiting reducts (decompose,
contract, refocus, contract, refocus, etc.)~is referred to as a
``reduction-free'' evaluation since it does not construct the successive
reducts.

There is evidence that such a refocus function exists and is useful,
witness the very existence of abstract machines since they directly
transition from actual redex to the next potential redex (or to the final
result), without detouring via the intermediate reduct.
Furthermore, if this refocus function could be expressed tail
recursively, it would provide the control flow of an abstract machine --
not by inventing this abstract machine, but by discovering it.

\subsection{From reduction-based evaluation to reduction-free evaluation}
\label{subsec:from-reduction-based-evaluation-to-reduction-free-evaluation}

In the early 2000's~\cite{Danvy-Nielsen:RULE01,Danvy-Nielsen:RS-04-26-shortest},
Nielsen and the author observed that the refocus function, essentially,
is the decomposition function itself, and they provided an informal proof
of this observation as well as a variety of examples of its relevance.
Since then, several application-specific proofs of refocusing have been
put forward (see \sectionref{sec:correctness-proofs-of-refocusing}) and
even more applications have been reported, starting with Biernacka and
the author's syntactic correspondence between calculi and abstract
machines~\cite{Biernacka-Danvy:TOCL07,Biernacka-Danvy:TCS07}.
The point of the present article is to present a simple formal proof that
is aligned with the simplicity of the original idea.

\subsection{Roadmap}
\label{subsec:roadmap}

\sectionref{sec:SOS-in-direct-style}
formalizes
a Structural Operational
Semantics~\cite{Plotkin:TR81-original,Plotkin:JLAP04} for the programming
language of discourse in Gallina, the resident pure and total functional
language of the Coq Proof Assistant.
This formalization is structurally recursive and in direct style.
\sectionref{sec:SOS-with-one-undelimited-continuation} presents an
equivalent formalization of this Structural Operational
Semantics that uses one undelimited continuation -- undelimited because
its codomain is polymorphic.
\sectionref{sec:SOS-with-one-undelimited-continuation-and-one-discontinuity}
presents an equivalent formalization of this Structural Operational
Semantics that uses one delimited continuation and one discontinuity --
delimited because its codomain is monomorphic, and with one discontinuity
because for stuck terms, the reduction function does not apply the
current continuation: it stops with an error message instead.
\sectionref{sec:SOS-with-two-continuations-and-one-discontinuity}
presents an equivalent formalization of this Structural Operational
Semantics that uses two delimited continuations -- one for
decomposing and one for recomposing -- and one
discontinuity for stuck terms.
\sectionref{sec:SOS-with-one-context-and-one-discontinuity}
presents a defunctionalized version of the formalization of
\sectionref{sec:SOS-with-two-continuations-and-one-discontinuity}.
This non-compositional format is independently known as that of a context
semantics~\cite{Felleisen:PhD} or again of a reduction
semantics~\cite{Felleisen-al:09}.
\sectionref{sec:SOS-with-one-undelimited-continuation-and-one-context-and-one-discontinuity}
presents an equivalent formalization of the Structural Operational
Semantics of
\sectionref{sec:SOS-with-two-continuations-and-one-discontinuity} that
uses one delimited continuation for decomposing, one context for
recomposing, and one discontinuity for stuck terms.
This formalization is compositional and therefore
expressible in Gallina.
Using this formalization, \sectionref{sec:the-simple-proof} presents a
simple proof of the refocusing theorem by induction on the context.
\sectionref{sec:assessment} assesses the optimality of refocusing as well
as its intrinsic and extrinsic applicability.
\sectionref{sec:related-work} reviews related work: the context where
refocusing arose, its formal correctness proofs,
and its many applications
\sectionref{sec:conclusion-and-perspectives} concludes and draws
perspectives.
\appendixref{app:defunctionalization} presents defunctionalization and
illustrates it with a fitting example.

\subsection{Prerequisites and notations}
\label{subsec:prerequisites-and-notations}

An elementary knowledge of Structural Operational Semantics, of Gallina,
and of the Coq Proof Assistant is expected from the reader.
The entirety of this work is formalized in
Coq
and all the propositions, lemmas, and theorem are proved in the
accompanying \inlinecoq{.v} file, all the way to the isomorphism
mentioned at the end of \appendixref{app:defunctionalization}.

\section{A Structural Operational Semantics, Functionally, In Direct Style}
\label{sec:SOS-in-direct-style}

Let us implement in Gallina the one-step reduction function of a
Structural Operational Semantics of the programming language of
discourse:
\begin{itemize}[leftmargin=3.5mm]

\item
  a literal reduces to a value;

\item
  if the term \inlinecoq{t1} is stuck, \\
  then so is the term \inlinecoq{Opr t1 opr t2}, for any term \inlinecoq{t2};

\item
  if the term \inlinecoq{t1} reduces to a term \inlinecoq{t1'}, \\
  then for any term \inlinecoq{t2},
  \inlinecoq{Opr t1 opr t2} reduces to \inlinecoq{Opr t1' opr t2};

\item
  if the term \inlinecoq{t1} reduces to a value \inlinecoq{v1},
  then

  \begin{itemize}[leftmargin=3.5mm]

  \item
    if the term \inlinecoq{t2} is stuck,
    then so is the term \inlinecoq{Opr t1 opr t2};

  \item
    if the term \inlinecoq{t2} reduces to a term \inlinecoq{t2'},
    then \inlinecoq{Opr t1 opr t2} reduces to \inlinecoq{Opr t1 opr t2'};

  \item
    if the term \inlinecoq{t2} reduces to a value \inlinecoq{v2},
    then \inlinecoq{Opr t1 opr t2} is a potential redex:
 
    \begin{itemize}[leftmargin=3.5mm]

    \item
      if this potential redex is an actual one,
      then contracting it yields a term; \\
      \inlinecoq{Opr t1 opr t2} reduces to this term;

    \item
      if this potential redex is not an actual one,
      then \inlinecoq{Opr t1 opr t2} is stuck.

    \end{itemize}
  \end{itemize}
\end{itemize}

\noindent
So reducing a term can yield
\begin{itemize}[leftmargin=3.5mm]

\item
  a value if this term does not contain a potential redex,

\item
  the next term in the reduction sequence if this term contains a
  potential redex which is an actual one (and therefore can be
  contracted), and

\item
  an error message if the term is stuck, \ie, contains a potential redex
  which is not an actual one, and therefore cannot be contracted:

\end{itemize}
\inputcoq{VALUE_OR_TERM_OR_STUCK}

\noindent
As befits a Structural Operational Semantics, the
implementation
is structurally recursive, and each of its recursive calls
emulates the traversal of the associated proof tree, which is constructed
depth first and from left to right.
In words --
\begin{itemize}[leftmargin=3.5mm]

\item
  reducing a given literal yields a value;

\item
  reducing a given operation means first attempting to reduce its left
  operand, which yields a value, a term, or an error message:

  \begin{itemize}[leftmargin=3.5mm]

  \item
    if reducing the given left operand yields an error message, then this
    left operand is stuck, and so is the given operation; reducing the
    given operation yields this error message;

  \item
    if reducing the given left operand yields a term, then this left operand
    contains an actual redex that was contracted; reducing the given
    operation yields an operation whose first operand is this term and
    whose right operand is the given right operand; and

  \item
    if reducing the given left operand yields a value, then there is
    nothing to reduce in the left operand, and so we attempt to reduce
    the given the given right operand, which yields a value, a term, or
    an error message:

    \begin{itemize}[leftmargin=3.5mm]

    \item
      if reducing the given right operand yields an error message, then this
      right operand is stuck, and so is the given operation; reducing the
      given operation yields this error message;

    \item
      if reducing the given right operand yields a term, then this right
      operand contains an actual redex that was contracted; reducing the
      given operation yields an operation whose first operand is
      the term representing the value obtained by reducing the given left operand
      (\ie, the given left operand itself)
      and whose right operand is this term; and

    \item
      if reducing the given right operand yields a value, then the given
      operation is a potential redex, which we attempt to contract:

      \begin{itemize}[leftmargin=3.5mm]

      \item
        if contraction succeeds, it yields a term which is the result of
        reducing the given operation; and

      \item
        if contraction fails, it yields an error message which is the
        result of reducing the given operation.

      \end{itemize}

    \end{itemize}

  \end{itemize}

\end{itemize}

\noindent
Concretely:
\inputcoq{TERM_OF_VALUE}
\inputcoq{REDUCE_D}

\noindent
(The accompanying \inlinecoq{.v} file contains a variation on
\inlinecoq{reduce_d} where \inlinecoq{Term (Opr t1 opr t2')} is returned
instead of \inlinecoq{Term (Opr (term_of_value v1) opr t2')} and a proof
of its equivalence.)

Evaluation is implemented as iterated reduction.
To make it structurally recursive and therefore a valid Gallina program,
we thread a decreasing counter in the manner of Bove and
Capretta~\cite{Bove-Capretta:MSCS05}:

\inputcoq{NORMALIZE_D}

\noindent
Each reduction step performs an operation -- and therefore removes one
operator -- in the given term.
So given a natural number that is larger than the number of operators in any
given term, normalizing this given term terminates:
\inputcoq{NORMALIZE_D_TERMINATES}

This implementation of one-step reduction is in direct style.
The next section describes an implementation in continuation-passing
style (CPS).

\section{A Structural Operational Semantics, Functionally, With One
  Undelimited Continuation}
\label{sec:SOS-with-one-undelimited-continuation}

As a first stepping stone towards refocusing,
here is
an equivalent implementation of the one-step reduction function that is
in (undelimited) continuation-passing style.
Again, as befits a Structural Operational Semantics, the implementation
is structurally recursive, and each of its tail-recursive calls emulates
the traversal of the associated proof tree, which is constructed depth
first and from left to right.
In words, given a continuation --
\begin{itemize}[leftmargin=3.5mm]

\item
  reducing a given literal continues with a value;

\item
  reducing a given operation means first attempting to reduce its left
  operand with a new continuation, which is then sent a value, a term, or
  an error message:

  \begin{itemize}[leftmargin=3.5mm]

  \item
    if the continuation is applied to an error message, then this
    left operand is stuck, and so is the given operation; reducing the
    given operation continues with this error message;

  \item
    if the continuation is applied to a term, then this left operand
    contains an actual redex that was contracted; reducing the given
    operation continues with an operation whose first operand is this term and
    whose right operand is the given right operand; and

  \item
    if the continuation is applied to a value, then there is nothing to
    reduce in the left operand, and so we attempt to reduce the given
    right operand with a new continuation, which is then sent a value, a
    term, or an error message:

    \begin{itemize}[leftmargin=3.5mm]

    \item
      if the continuation is applied to an error message, then this
      right operand is stuck, and so is the given operation; reducing the
      given operation continues with this error message;

    \item
      if the continuation is applied to a term, then this right
      operand contains an actual redex that was contracted; reducing the
      given operation continues with an operation whose first operand is
      the given left operand
      and whose right operand is this term; and

    \item
      if the continuation is applied to a value, then the given
      operation is a potential redex, which we attempt to contract:

      \begin{itemize}[leftmargin=3.5mm]

      \item
        if contraction succeeds, it yields a term and reducing the given
        operation continues with this term; and

      \item
        if contraction fails, it yields an error message and reducing the given
        operation continues with this error message.

      \end{itemize}

    \end{itemize}

  \end{itemize}

\end{itemize}

\noindent
Concretely:
\inputcoq{REDUCE3_CP}
\inputcoq{REDUCE3_C}

\noindent
The codomain of \inlinecoq{reduce3_c'} is polymorphically typed and so is
the codomain of its continuation, which is thus undelimited.
This tail-recursive implementation of one-step reduction is in
continuation-passing style~\cite{Strachey-Wadsworth:TR74-original-shorter,Thielecke:ESOP04,Wadsworth:HOSC00}.
\inputcoq{REDUCE_D_AND_REDUCE3_C_ARE_EQUIVALENT}

Analyzing the structure of the continuation, we see that when the given
term is stuck, the continuation is applied to an error message, which
triggers the previous continuation (\inlinecoq{k3}) to be applied to an
error message, etc., all the way to the initial continuation, just as in
the definition of \inlinecoq{reduce_d}.
If we were to split the continuation into a pair of continuations, one
applied to values or to terms, and the other to error messages, we could
bypass this propagation of error messages.
(This splitting is made possible by the isomorphism between $A + B
\rightarrow C$ and $(A \rightarrow C) \times (B \rightarrow C)$.)
Furthermore, we would then be in position to anticipate the instantiation of
the polymorphic type of answers in the definition of
\inlinecoq{reduce3_c} and directly apply the second continuation for
potential redexes that are not actual ones to return the error message.
This massaging leads to the one-step reduction function described in the
next section, where
\begin{itemize}

\item
  the continuation is delimited because its codomain of answers is
  monomorphically typed, and

\item
  for potential redexes that are not actual ones, the continuation is not
  applied, which is a discontinuity.

\end{itemize}

\section{A Structural Operational Semantics, Functionally, With One
  Delimited Continuation And One Discontinuity}
\label{sec:SOS-with-one-undelimited-continuation-and-one-discontinuity}

As a second stepping stone towards refocusing,
here is
an equivalent implementation of the one-step reduction function that is
in delimited continuation-passing style and has a discontinuity.
As still befits a Structural Operational Semantics, the implementation
is structurally recursive, and each of its tail-recursive calls emulates
the traversal of the associated proof tree, which is constructed depth
first and from left to right.
In words, given a continuation --
\begin{itemize}[leftmargin=3.5mm]

\item
  reducing a given literal continues with a value;

\item
  reducing a given operation means first attempting to reduce its left
  operand with a new continuation, which is then sent a value or a term:

  \begin{itemize}[leftmargin=3.5mm]

  \item
    if the continuation is applied to a term, then this left operand
    contains an actual redex that was contracted; reducing the given
    operation continues with an operation whose first operand is this term and
    whose right operand is the given right operand; and

  \item
    if the continuation is applied to a value, then there is nothing to
    reduce in the left operand, and so we attempt to reduce the given
    right operand with a new continuation, which is then sent a value or
    a term:

    \begin{itemize}[leftmargin=3.5mm]

    \item
      if the continuation is applied to a term, then this right
      operand contains an actual redex that was contracted; reducing the
      given operation continues with an operation whose first operand is
      the given left operand
      and whose right operand is this term; and

    \item
      if the continuation is applied to a value, then the given
      operation is a potential redex, which we attempt to contract:

      \begin{itemize}[leftmargin=3.5mm]

      \item
        if contraction succeeds, it yields a term and reducing the given
        operation continues with this term; and

      \item
        if contraction fails, it yields an error message and reducing the given
        operation directly yields this error message, which is the discontinuity.

      \end{itemize}

    \end{itemize}

  \end{itemize}

\end{itemize}

\noindent
Concretely:
\inputcoq{VALUE_OR_TERM}
\inputcoq{REDUCE2_CP}
\inputcoq{REDUCE2_C}

\noindent
The codomain of \inlinecoq{reduce2_c'} is monomorphically typed and so is
the codomain of its continuation, which is thus
delimited~\cite{Danvy-Filinski:LFP90}.
So stricto sensu this tail-recursive implementation of one-step reduction
is not in continuation-passing
style --
it is in delimited continuation-passing style.
\inputcoq{REDUCE_D_AND_REDUCE2_C_ARE_EQUIVALENT}

Analyzing the structure of the continuation, we see that
either it continues the decomposition or it carries out a recomposition.
In the next section, we split this delimited continuation into two: one
delimited continuation that is applied to values to continue the
decomposition, and one delimited continuation that is applied to terms to
carry out the recomposition.

\section{A Structural Operational Semantics, Functionally, With Two
  Delimited Continuations And One Discontinuity}
\label{sec:SOS-with-two-continuations-and-one-discontinuity}

As a third stepping stone towards refocusing,
here is
an equivalent implementation of the one-step reduction function that uses
two delimited continuations -- one that is applied to values to continue
the decomposition and has type \inlinecoq{value -> value_or_term}, and
one that is applied to terms to carry out the recomposition and has type
\inlinecoq{term -> term} -- and one discontinuity
for stopping
with an error message in case the potential redex is not an actual one.
\inputcoq{VALUE_OR_DECOMPOSITION_KK}

\noindent
We refer to the continuation of type \inlinecoq{value -> value_or_term}
as ``the decomposing continuation'' and the other, which has type
\inlinecoq{term -> term}, as ``the recomposing continuation.''

\subsection{Decomposition}
\label{subsec:SOS-with-two-continuations-and-one-discontinuity:decomposition}

Let us factor out the decomposition facet of the one-step reduction
function into a (tail-recursive) decomposition function:
\inputcoq{DECOMPOSE_TERM_KK}
\inputcoq{DECOMPOSE_KK}

This decomposition function either returns a value or a decomposition.
In the latter case, this decomposition contains a potential redex and the
two continuations that were in effect when this potential redex was
found.

As mentioned in
\sectionref{subsec:small-step-semantics-of-the-programming-language-of-discourse},
the recomposing continuation is a left inverse of the decomposition function:
\inputcoq{TERM_OF_POTENTIAL_REDEX}
\inputcoq{THERE_AND_BACK_KK}

\subsection{One-step reduction}
\label{subsec:SOS-with-two-continuations-and-one-discontinuity:one-step-reduction}

Using the decomposition function, the one-step reduction function is
stated as follows:
\inputcoq{REDUCE_KK}

\noindent
To prove its equivalence with the one-step reduction function in direct
style (\sectionref{sec:SOS-in-direct-style}), we make use of the
following three lemmas that relate \inlinecoq{reduce_d} and
\inlinecoq{decompose_term_cc}.
\begin{itemize}[leftmargin=3.5mm]

\item
  For any given term, if applying \inlinecoq{reduce_d} to it yields a
  value, then applying \inlinecoq{decompose_term_cc} to this term, a
  decomposing continuation, and a recomposing continuation yields
  the same result as applying this decomposing continuation to this
  value:
  \inputcoq{ABOUT_REDUCE_D_AND_DECOMPOSE_TERM_KK_VALUE}

\item
  For any given term, if applying \inlinecoq{reduce_d} to it yields
  another term, then (1) applying \inlinecoq{decompose_}
  \inlinecoq{term_cc} to this term, a decomposing continuation, and a
  recomposing continuation yields a decomposition of this term into a
  potential redex, another decomposing continuation, and another
  recomposing continuation, (2) contracting the potential redex yields a
  contractum and (3) applying the new recomposing continuation to this
  contractum yields the same reduct as applying the recomposing
  continuation to this other term:
  \inputcoq{ABOUT_REDUCE_D_AND_DECOMPOSE_TERM_KK_TERM}

\item
  For any given term, if applying \inlinecoq{reduce_d} to it yields an
  error message, then applying \inlinecoq{decompose_term_cc} to this
  term, a decomposing continuation, and a recomposing continuation
  yields a potential redex whose contraction yields the same error
  message:
  \inputcoq{ABOUT_REDUCE_D_AND_DECOMPOSE_TERM_KK_STUCK}

\end{itemize}

\noindent
The equivalence of \inlinecoq{reduce_d} and \inlinecoq{reduce_cc}
follows, and likewise for the corresponding reduction-based normalization
function:
\inputcoq{REDUCE_D_AND_REDUCE_KK_ARE_EQUIVALENT}
\inputcoq{NORMALIZE_KK}
\inputcoq{NORMALIZE_D_AND_NORMALIZE_KK_ARE_EQUIVALENT}

\subsection{A more convenient reduction-based normalization function}
\label{subsec:a-more-convenient-reduction-based-normalization-function}

Currently, the normalization function tests the result of one-step
reduction and the reduction function tests the result of the
decomposition function.
Let us streamline these two tests into one by iterating on the result of
decomposition, \ie, the final value or the next decomposition:
\inputcoq{ITERATE_KK_RB}
\vspace{-1mm}
\inputcoq{NORMALIZE_KK_RB}

We are getting close to the first diagram of
\sectionref{subsec:small-step-semantics-of-the-programming-language-of-discourse}.
(The slanted down arrow is the recomposing continuation.)
In words -- after an initial decomposition, the iteration function tests
whether normalization should stop (with a value or with an error message)
or continue by (1) recomposing the reduction context over the contractum
with \inlinecoq{kr} and (2) decomposing the resulting reduct with
\inlinecoq{decompose_cc}, as underlined above with {\small{\verb"(* ^^^ *)"}}.
\inputcoq{NORMALIZE_D_AND_NORMALIZE_KK_RB_ARE_EQUIVALENT}

\vspace{-1mm}

\subsection{The corresponding reduction-free normalization function}

The thesis here is that there is no need to recompose the context
around the contractum and then decompose the resulting reduct -- one can
instead continue the decomposition over the contractum to obtain a final
value or the next decomposition with \inlinecoq{decompose_term_cc}, as
underlined below with {\small{\verb"(* ^^^ *)"}}.
\inputcoq{ITERATE_KK_RF}
\vspace{-1mm}
\inputcoq{NORMALIZE_KK_RF}

We are getting close to the second diagram of
\sectionref{subsec:small-step-semantics-of-the-programming-language-of-discourse}.
(The dashed arrow (\ie, refocus) is the decomposition function.)

Here is the equivalence between reduction-based normalization and
reduction-free normalization we want to prove:
\inputcoq{ITERATE_KK_RB_AND_ITERATE_KK_RF_ARE_EQUIVALENT}

By extension, here is the refocusing theorem we want to prove:
\inputcoq{WANTED}

\noindent
In words -- if decomposing a given term yields a potential redex, a
decomposition continuation, and a recomposing continuation, then for any
term, it is equivalent (a) to recompose the reduction context over this
term and then decompose the result and (b) to continue the decomposition
of this term with the decomposition continuation and the recomposing
continuation.

Proving this theorem would most naturally be done by induction over the
recomposing continuation, were it not for it being a function.
So, defunctionalization to the rescue.
In \sectionref{sec:SOS-with-one-context-and-one-discontinuity}, we
defunctionalize both the decomposing continuation and the recomposing
continuation, but even though the result is of independent interest, that
is overshooting.
So in
\sectionref{sec:SOS-with-one-undelimited-continuation-and-one-context-and-one-discontinuity},
we only defunctionalize the recomposing continuation, and that is
enough for proving the refocusing theorem by induction on the data type
of the recomposing continuation -- \ie, the context -- in
\sectionref{sec:the-simple-proof}.

\section{A Structural Operational Semantics, Functionally, With One
  Context And One Discontinuity}
\label{sec:SOS-with-one-context-and-one-discontinuity}

Since they were split from the same delimited continuation in
\sectionref{sec:SOS-with-one-undelimited-continuation-and-one-discontinuity},
defunctionalizing the two delimited continuations in
\sectionref{sec:SOS-with-two-continuations-and-one-discontinuity} yields
the same data type of reduction contexts and two dispatching functions:
one to continue the decomposition and one to continue the recomposition.
This format is independently known as that of a context
semantics~\cite{Felleisen:PhD} or again of a reduction
semantics~\cite{Felleisen-al:09}.
Implementing it in Gallina, however, is
not immediate because -- as usual --
after defunctionalizing the continuation of a recursive function, the
resulting program is not compositional.
Therefore, a functional representation requires an encoding in the manner
of Bove and Capretta~\cite{Bove-Capretta:MSCS05} to make it
structurally recursive.
Alternatively, the representation could be relational instead of
functional, which
brings us from the frying pan into two fires since
the formalization is no longer an executable functional program and
unique decomposition becomes an issue that needs to be addressed.

The next section goes halfway by only defunctionalizing the recomposing
continuation: the result is still compositional and therefore expressible
in Gallina.

\section{A Structural Operational Semantics, Functionally, With One
  Delimited Continuation, One Context, And One Discontinuity}
\label{sec:SOS-with-one-undelimited-continuation-and-one-context-and-one-discontinuity}

In this section, we only defunctionalize the recomposing continuation
from \sectionref{sec:SOS-with-two-continuations-and-one-discontinuity},
which gives rise to a grammar of recomposing contexts.
These contexts can be represented inside-out
(\sectionref{subsec:inside-out-contexts}) or outside-in
(\sectionref{subsec:outside-in-contexts}).
We then implement the decomposition function with a decomposing
continuation, a recomposing context, and a discontinuity, and we prove
that the recomposition function is a left inverse of the decomposition
function
(\sectionref{subsec:SOS-with-one-undelimited-continuation-and-one-context-and-one-discontinuity:decomposition}).
In addition though, we present an inductive relation that connects the
decomposing continuation and the recomposing context, and we prove that
if decomposing a term yields a potential redex, a decomposing
continuation, and a recomposing context, then this continuation and this
context are in this relation.
We then implement the corresponding one-step reduction function
(\sectionref{subsec:SOS-with-one-undelimited-continuation-and-one-context-and-one-discontinuity:one-step-reduction}),
the corresponding reduction-based normalization function
(\sectionref{subsec:the-reduction-based-normalization-function}), and the
corresponding reduction-free normalization function
(\sectionref{subsec:the-reduction-free-normalization-function}).
And that puts us in position to prove the refocusing theorem by
induction over the context (\sectionref{sec:the-simple-proof}).

\subsection{Inside-out contexts}
\label{subsec:inside-out-contexts}

Let us defunctionalize the recomposing continuation.
The resulting data type is that of the inside-out contexts from the
left-to-right reduction strategy,
which independently gave rise to `zippers'~\cite{Huet:JFP97}.
This data type is flat and so isomorphic to lists.
Accordingly, let us represent a context as a list of control frames:
\inputcoq{CONTROL_FRAME}
\inputcoq{CONTEXT}

Since the context is inside-out and was accumulated iteratively by the
decomposition function, its associated recomposition function is also
tail recursive:
\inputcoq{RECOMPOSE_IO}

\subsection{Outside-in contexts}
\label{subsec:outside-in-contexts}

Nothing prevents one from representing contexts outside in.
(To quote, ``contexts are specified so that a context can easily be
extended to the left or to the
right~\cite[Section~4.3]{Danvy-Nielsen:RS-01-23-short}.)
The associated recomposition function is then recursive:
\inputcoq{RECOMPOSE_OI}

The two recomposition functions are related in the manner of Burge and
Landin's loop fusion~\cite[page~114]{Burge:75} and of Bird and
Wadler's third duality theorem~\cite[page~68]{Bird-Wadler:88}:
\inputcoq{ABOUT_RECOMPOSING_IO_AND_RECOMPOSING_OI}

\noindent
In words -- outside-in contexts and inside-out contexts are reverse of
each other.

The accompanying \inlinecoq{.v} file contains a recursive implementation
of decomposition that returns a value or a potential redex together with
its context.
This context is not accumulated at call time (like the accumulator in the
tail-recursive list reverse function), it is constructed at return time
(like the result of the recursive list copy, list append, and list map
functions).
So if the tail-recursive decomposition function yields a potential redex
and its context, this context is inside out, and if the recursive
decomposition function yields a potential redex and its context, this
context is outside in.
As proved in the accompanying \inlinecoq{.v}, these two potential redexes
are the same one and these two contexts are inverses of each other.

\subsection{Decomposition}
\label{subsec:SOS-with-one-undelimited-continuation-and-one-context-and-one-discontinuity:decomposition}

Here is the defunctionalized counterpart of the result of decomposition
in \sectionref{sec:SOS-with-two-continuations-and-one-discontinuity}:
\inputcoq{VALUE_OR_DECOMPOSITION_KC}

Here is the defunctionalized counterpart of the decomposition function
in \sectionref{subsec:SOS-with-two-continuations-and-one-discontinuity:decomposition}
\inputcoq{DECOMPOSE_TERM_KC}
\inputcoq{DECOMPOSE_KC}

This decomposition function either returns a value or a decomposition,
and this decomposition contains a potential redex and the decomposing
continuation and recomposing context that were in effect when this
potential redex was found.

As mentioned in
\sectionref{subsec:small-step-semantics-of-the-programming-language-of-discourse},
the recomposition function is a left inverse of the decomposition function:
\inputcoq{THERE_AND_BACK_KC}

The following relation characterizes the correspondence between the
decomposing continuation and the recomposing context in a decomposition:
\inputcoq{CORR}

If decomposing a term yields a potential redex, a continuation, and a
context, then this continuation and this context are in this relation:
\inputcoq{DECOMPOSE_KC_YIELDS_CORR}

Incidentally, since defunctionalizing the decomposing continuation gives
rise to this very data type of contexts, this relation also connects this
decomposing continuation and its defunctionalized counterpart: it is the
relational content of defunctionalization (and refunctionalization) here.

\clearpage

\subsection{One-step reduction}
\label{subsec:SOS-with-one-undelimited-continuation-and-one-context-and-one-discontinuity:one-step-reduction}

Using the decomposition function, the one-step reduction function is
stated as follows:
\inputcoq{REDUCE_KC}

For brevity, we omit the
lemmas that relate \inlinecoq{reduce_d} and
\inlinecoq{decompose_term_cC}.
Their statements can be deduced from the similar lemmas in
\sectionref{subsec:SOS-with-two-continuations-and-one-discontinuity:one-step-reduction}.
(The accompanying \inlinecoq{.v} file contains them as well as their proof.)

The equivalence of \inlinecoq{reduce_d} and \inlinecoq{reduce_cC}
follows:
\inputcoq{REDUCE_D_AND_REDUCE_KC_ARE_EQUIVALENT}

\subsection{The reduction-based normalization function}
\label{subsec:the-reduction-based-normalization-function}

Here is the defunctionalized counterpart of the reduction-based
normalization function from
\sectionref{subsec:a-more-convenient-reduction-based-normalization-function}:
\inputcoq{ITERATE_KC_RB}
\inputcoq{NORMALIZE_KC_RB}

We are now implementing the first diagram of
\sectionref{subsec:small-step-semantics-of-the-programming-language-of-discourse}.
\inputcoq{NORMALIZE_D_AND_NORMALIZE_KC_RB_ARE_EQUIVALENT}

\subsection{The reduction-free normalization function}
\label{subsec:the-reduction-free-normalization-function}

The thesis here is still that there is no need to recompose the context
around the contractum and then decompose the resulting reduct -- one can
instead continue the decomposition over the contractum to obtain a final
value or the next decomposition:

\ 

\inputcoq{ITERATE_KC_RF}
\inputcoq{NORMALIZE_KC_RF}

We are now implementing the second diagram of
\sectionref{subsec:small-step-semantics-of-the-programming-language-of-discourse}.

Here is the equivalence between reduction-based normalization and
reduction-free normalization
we wish for,
and that we are now in position to prove simply:
\inputcoq{ITERATE_KC_RB_AND_ITERATE_KC_RF_ARE_EQUIVALENT}

\vspace{-2mm}

\section{The Simple Proof}
\label{sec:the-simple-proof}

After de Bruijn presented his eponymous indices and levels to represent
bound variables in $\lambda$ terms~\cite{de-Bruijn:IM72}, the indices
became a hit because they correspond to the lexical offsets of the bound
variables~\cite{Wand:IPL90}.
Later in his life, de Bruijn pointed out that the levels are a better fit when
generating $\lambda$ terms, which is very true in normalization by
evaluation for example~\cite{Filinski:PPDP99}.
A similar situation occurs here.
A priori we want to write something like the following:
\inputcoq{REFOCUSING}

\noindent
In words -- given a term that can be decomposed into a potential redex, its
decomposing continuation, and its recomposing context, decomposing
the result of recomposing this context over any other term gives the
same result as decomposing this other term with the same context and the
same continuation.

Proving this theorem is challenging because \inlinecoq{decompose_cC}
consumes the outer component of the recomposed term, which is constructed
last since the context is inside out.
If the context were outside in, we would be in position to ``cancel out''
the outside constructors of the context and their subsequent
decomposition, inductively, in the proof.

So should we represent contexts outside in?
That would inflict a linear overhead to extending the contexts during the
decomposition.
Instead, let us exploit
the relation between \inlinecoq{recomposing_io} and \inlinecoq{recomposing_oi}
from \sectionref{subsec:outside-in-contexts} and transiently reverse the
context once it has been constructed -- in the proof, not in the program.

Generalizing, here is a Eureka lemma that captures what happens in the
middle:
\inputcoq{REFOCUSING_AUX}

\noindent
In words -- we are given a term, a potential redex, a decomposing
continuation \inlinecoq{k_rf}, an outside-in context, and an inside-out
context that are such that the term decomposes into this potential redex,
\inlinecoq{k_rf}, and the concatenation of the reverse of the outside-in
context and the inside-out context.
Then, for any decomposing continuation \inlinecoq{k_rb} that is related
to the inside-out context, decomposing the result of recomposing the
outside-in context over any other term with \inlinecoq{k_rb} and the
inside-out context gives the same result as decomposing this other term
with \inlinecoq{k_rf} and the concatenated context.

This lemma is proved by induction on \inlinecoq{C_io}.

The refocusing theorem is a corollary of this lemma, witness the last
line of its proof:
\inputcoq{REFOCUSING_WITH_A_STRIPPED_PROOF}

The equivalence of \inlinecoq{iterate_cC_rb} and
\inlinecoq{iterate_cC_rf} follows, and so does the equivalence of
\inlinecoq{normalize_cC_rb} and \inlinecoq{normalize_cC_rf}.

\section{Assessment}
\label{sec:assessment}

This section briefly
discusses the optimality of the refocus function,
analyzes the applicability of refocusing,
and
puts this applicability into perspective.

\subsection{Optimality of the traversal after refocusing}

If there is something to optimize after refocusing then it is the
decomposition function that needs to be optimized, and that means that it
was not optimal in the first place, which is hard to imagine for the
functional implementation of the reduction strategy in a Structural
Operational Semantics when this strategy is deterministic.
Here is another argument for this optimality:
In the syntactic correspondence between calculi and abstract
machines~\cite{Biernacka-Danvy:TCS07}, the next step after refocusing is
lightweight fusion by fixed-point promotion~\cite{Ohori-Sasano:POPL07} of
\inlinecoq{iterate_cC_rf} and \inlinecoq{decompose_term_cC}.
Performing this next step and inlining the contraction function then
yields the compositional evaluation function for arithmetic expressions
that one would authoritatively write by hand using a delimited
continuation and with a discontinuity to account for numerical underflow.
And likewise, defunctionalizing this continuation yields the canonical
abstract machine for arithmetic expressions with left-to-right evaluation
that is such a pleasure to invent~\cite{Danvy:AFP08}.
This compositional evaluation function and this abstract machine,
however, were not invented -- they were discovered.
(Precedents include
Knuth discovering and Morris inventing the KMP
string-matching algorithm~\cite{Knuth-Morris-Pratt:SIAM77}, 
Reynolds discovering~\cite{Reynolds:72-original}
and Felleisen and Friedman inventing~\cite{Felleisen-Friedman:FDPC3}
the CEK machine,
and
Schmidt discovering~\cite{Schmidt:HOSC07}
and Krivine designing~\cite{Krivine:HOSC07}
the Krivine machine --
three
illustrations of Rota's insightful piece about theorizers and problem
solvers~\cite[pages~45-46]{Rota:96}.)

\subsection{Applicability}

Refocusing scales to context-sensitive contraction
rules~\cite{Biernacka-Danvy:TCS07}.

For outermost reduction strategies~\cite{Danvy-Johannsen:LOPSTR13}, if
the contraction rules overlap, one needs to recompose the contractum with
a prefix of the context to resolve the ambiguity -- a local solution for
a local problem -- and then the decomposition can unambiguously be
continued.

The salient point here is that we are not inventing, we are discovering
and ending with the reduction-free normalization function that one would
-- a priori or a posteriori -- happily invent in the first place.
The author is aware that the previous sentence contains a strong
statement -- as well as a subjective one -- but this statement has not been
invalidated in the 20+ years of existence of refocusing.

\subsection{On the inter-derivation of semantic artifacts}

Just like the measure of goodness for a programming language is how
convenient a notation it offers for expressing computation (\ie, for
programming), the measure of goodness for a formal semantics is how
convenient a notation it offers for reasoning (\ie, for proving).
In practice -- and the present article makes no exception -- formal
semantics are inter-derived to put oneself in position to prove a
theorem.
In the present case,
\sectionsto{sec:SOS-in-direct-style}{sec:SOS-with-one-undelimited-continuation-and-one-context-and-one-discontinuity}
document such a derivational journey.
Consider the derived Structural Operational Semantics with one delimited
continuation, one context, and one discontinuity
at the end of this journey.
A decomposition is a triple -- potential redex, recomposing context,
and decomposing continuation.
The reduction-based normalization function
(\sectionref{subsec:the-reduction-based-normalization-function}) ignores
the decomposing continuation and the reduction-free normalization
function (\sectionref{subsec:the-reduction-free-normalization-function})
ignores the recomposing context.
And so this semantics is teetering between reduction-based normalization
and reduction-free normalization,
making it simple
to state the refocusing theorem and prove it.

More broadly, though, all the work invested in inter-deriving formal
semantics can also be seen as a collective effort for establishing the
semantic analogue of Turing completeness for programming languages, which
makes some kind of sense considering the Curry-Howard correspondence,
where programming is proving and proving is programming.
Induction proofs would be akin to recursive programming, rule induction to
logic programming, coinduction to rewriting, etc., suggesting which kind
of program should be extracted from which kind of proof.

To sketch a taxonomy for inter-deriving semantic artifacts based on the
operational content of their format,
\begin{itemize}[leftmargin=3.5mm]

\item
  refocusing connects reduction-based normalization
  (discrete, observable reduction steps) and
  reduction-free normalization (continuous, unobservable reduction steps),

\item
  lightweight fusion by fixed-point promotion~\cite{Ohori-Sasano:POPL07}
  connects small-step normalization and big-step
  normalization~\cite{Danvy-Millikin:IPL08},

\item
  continuation-passing
  style~\cite{Sussman-Steele:75-original,Sussman-Steele:HOSC98-revisited}
  ensures evaluation-order
  independence~\cite{Plotkin:TCS75,Reynolds:72-original,Reynolds:HOSC98-revisited},
  in one big step,

\item
  direct style provides a granularity for big steps,

\item
  defunctionalization~\cite{Reynolds:72-original,Reynolds:HOSC98-revisited}
  provides first-orderness, witness the rational reconstruction of
  reduction semantics in
  \sectionref{sec:SOS-with-one-context-and-one-discontinuity},

\item
  closure conversion~\cite{Landin:CJ64} connects compositional and
  non-compositional normalization functions when expressible values are
  functions,

\item
  explicit substitutions~\cite{Abadi-al:JFP91,Curien:TCS91} provide a
  foundation for environments~\cite{Biernacka-Danvy:TOCL07,Landin:CJ64}, and

\item
  abstract machines provide a welcoming intersection where theoreticians
  feel that they are doing implementation work and implementors feel
  that they are doing theory.

\end{itemize}

\noindent
The elements above do not fall out of the sky -- they are part of the
syntactic correspondence between context-sensitive calculi and abstract
machines~\cite{Biernacka-Danvy:TCS07} and of the functional
correspondence between evaluators and abstract
machines~\cite{Ager-al:TCS05}, but the taxonomy above is just a sketch.
Still, this sketch seems to be in the right ballpark, considering the
coincidence of the reduction contexts in a small-step semantics and of
the evaluation contexts in a big-step semantics.
The reduction contexts in a small-step semantics are interpreted by the
recomposition function, and are the defunctionalized counterpart of the
continuation in the corresponding one-step reduction function.
The evaluation contexts in a big-step semantics are interpreted by a
``continue'' function (cf.\ eval/continue abstract machines) and are the
defunctionalized counterpart of the continuation in the corresponding
big-step evaluation function.
This coincidence heralds a global unity between small-step and big-step
semantics, one that links Plotkin's motto that well-typed programs do not
get stuck in a small-step semantics and Milner's motto that well-typed
programs do not go wrong in a big-step semantics as two sides of the same
computational coin.
Along the same line, it is also spectacular how monadic effects in a
big-step semantics~\cite{Ager-al:TCS05} are mirrored by context-sensitive
contraction rules in small-step semantics~\cite{Biernacka-Danvy:TCS07}.
So, ballpark.

\section{Related Work}
\label{sec:related-work}

This section reviews related work: the context where refocusing arose,
its few other correctness proofs, and its many applications.

\subsection{Context of refocusing}

Let us revisit the programming language of discourse and its small-step
semantics
(\sectionref{subsec:small-step-semantics-of-the-programming-language-of-discourse}).

Structural Operational Semantics is a small-step direct semantics and one
reduction step gives rise to constructing a proof tree.
At the top of this proof tree, a redex is contracted:
$$
\displaystyle
\frac
  {\displaystyle
   \frac
     {\displaystyle
      {\phantom{()}}
      \frac
       {}
       {5 + 5 \rightarrow 10}}
     {1 - (5 + 5) \rightarrow 1 - 10}}
  {(1 - (5 + 5)) - (2 - 20) \rightarrow (1 - 10) - (2 - 20)}
$$

\noindent
Let us reformat this proof tree to emphasize its implicit decomposition,
its contraction, and its implicit recomposition:
{\newcommand{\outercontext}[1]{{#1} - (2 - 20)}
 \newcommand{\outercontextp}[1]{\outercontext{({#1})}}
 \newcommand{\phantomoutercontext}[1]{{#1} \phantom{- (2 - 20)}}
 \newcommand{\phantomoutercontextp}[1]{\phantomoutercontext{\phantom{(}{#1}\phantom{)}}}
 \newcommand{\innercontext}[1]{1 - {#1}}
 \newcommand{\innercontextp}[1]{\innercontext{(#1)}}
 \newcommand{\phantominnercontext}[1]{\phantom{1 -} {#1}}
 \newcommand{\phantominnercontextp}[1]{\phantominnercontext{\phantom{(}{#1}\phantom{)}}}
 \let\labelstyle=\textstyle
 \spreaddiagramrows{-0.8cm}
 \spreaddiagramcolumns{-0.8cm}
 \newcommand{\atadless}{\hspace{-1mm}}
 $$
 \diagram
 \ 
 \rrto^{\Text{contraction}}
 &
 \ 
 &
 \ddto^{\Text{implicit \\ recomposition}}
 \\
 \ 
 &
 {\begin{array}{l@{\ }c@{\ }l}
    \phantomoutercontextp{\phantominnercontextp{\phantom{(}5 + 5}}
    &
    \rightarrow
    &
    \phantomoutercontextp{\phantominnercontextp{10}}
    \\
    \hline
    \phantomoutercontextp{\innercontextp{5 + 5}}
    &
    \rightarrow
    &
    \phantomoutercontextp{\innercontext{10}}
    \\
    \hline
    \outercontextp{\innercontextp{5 + 5}}
    &
    \rightarrow
    &
    \outercontextp{\innercontext{10}}
  \end{array}}
 &
 \ 
 \\
 \ 
 \uuto^{\Text{implicit \\ decomposition}}
 &
 \ 
 & 
 \enddiagram
 $$
}

Reduction semantics is a small-step continuation semantics where the
continuation is delimited and represented as a context, and so one
reduction step gives rise to (1) constructing the context of the redex,
(2) contracting the redex, and (3) recomposing the context around the
contractum to yield the reduct:
\begin{align*}
  \begin{array}{@{}l@{ }c}%
    \left.
    \begin{array}{@{}l@{ }}%
      \phantom{[}(1 - (5 + 5)) - (2 - 20)\phantom{]}
      \\
      {}[(1 - (5 + 5)) - (2 - 20)]
      \\
      {}[\phantom{\hspace{0.4mm}}[1 - (5 + 5)]\phantom{\hspace{0.4mm}} - (2 - 20)]
      \\
      {}[\phantom{\hspace{0.4mm}}[1 - \phantom{\hspace{0.4mm}}[5 + 5]\phantom{\hspace{0.4mm}}]\phantom{\hspace{0.4mm}} - (2 - 20)]
    \end{array}
    \right\rbrace
    &
    \begin{array}{@{}c@{}}
      \mathrm{explicit}
      \\
      \mathrm{decomposition}
    \end{array}
    \\
    &
    \mathrm{contraction}
    \\
    \left.
    \begin{array}{@{}l@{ }}%
    \phantom{\hspace{0.5mm}}[\phantom{\hspace{0.4mm}}[1 - \phantom{\hspace{0.4mm}}[\phantom{\hspace{2.26mm}}10\phantom{\hspace{2.26mm}}]\phantom{\hspace{0.4mm}}]\phantom{\hspace{0.4mm}} - (2 - 20)]
    \\
    {}[\phantom{\hspace{0.4mm}}[1 - \phantom{\hspace{0.4mm}}{\phantom{[}}\phantom{\hspace{2.28mm}}10\phantom{\hspace{2.28mm}}{\phantom{]}}\phantom{\hspace{0.4mm}}]\phantom{\hspace{0.4mm}} - (2 - 20)]
    \\
    {}[(1 - \phantom{\hspace{0.4mm}}{\phantom{[}}\phantom{\hspace{2.28mm}}10\phantom{\hspace{3.4mm}})\phantom{\hspace{0.4mm}} - (2 - 20)]
    \\
    {}\phantom{[}(1 - \phantom{\hspace{0.4mm}}{\phantom{[}}\phantom{\hspace{2.28mm}}10\phantom{\hspace{3.4mm}})\phantom{\hspace{0.4mm}} - (2 - 20)\phantom{]}
    \end{array}
    \right\rbrace
    &
    \begin{array}{@{}c@{}}
      \mathrm{explicit}
      \\
      \mathrm{recomposition}
    \end{array}
  \end{array}
\end{align*}

At the turn of the century~\cite{Xiao-al:RULE00,Xiao-al:HOSC01}, the
question arose how to ``short-circuit'' the construction of the
intermediate reduct to go directly from a contractum and its context to
the next potential redex and its context.
Refocusing is the answer to this question, and the value of this answer
is that as it happens, refocused reduction semantics are in one-to-one
correspondence with abstract machines that were independently
designed~\cite{Biernacka-Danvy:TOCL07,Biernacka-Danvy:TCS07}.

\subsection{Correctness proofs of refocusing}
\label{sec:correctness-proofs-of-refocusing}

Nielsen and the author stated structural conditions
about reduction semantics under which refocusing is possible, and
provided an informal proof for them~\cite{Danvy-Nielsen:RS-04-26-shortest}.
These structural conditions are essentially the same as in Xiao et al.'s
work~\cite{Xiao-al:RULE00, Xiao-al:HOSC01}.
In his MSc thesis~\cite{Sieczkowski:MS}, Sieczkowski formalized
refocusing using the Coq Proof Assistant.
Sieczkowski, Biernacka, and Biernacki presented a generic formalization
of refocusing and formally proved its correct, using the Coq
Proof Assistant~\cite{Sieczkowski-al:IFL10}.
In his PhD thesis~\cite{Bach-Poulsen:PhD}, Bach Poulsen proved the
correctness of refocusing for XSOS using rule induction.
When studying how the syntactic correspondence fares for outermost
reduction strategies instead of for innermost
ones~\cite{Danvy-Johannsen:LOPSTR13}, Johannsen and the author noticed
another condition about the contraction rules when they are
context-sensitive and overlap: after a contraction, decomposition should
not continue on the contractum with its context but on the partially
recomposed contractum that disambiguates the overlap and with the rest of
this context.
Garc{\'{\i}}a{-}P{\'{e}}rez and Nogueira characterized this other
condition as a hybrid reduction strategy~\cite{Garcia-Perez-Nogueira:SCP14}.
Biernacka, Charatonik, and Zielinska generalized refocusing to both
uniform and hybrid strategies and formally proved it correct, using the
Coq Proof Assistant~\cite{Biernacka-al:FSCD17-official}.
Using Agda, Swiestra formalized the syntactic correspondence between the
$\lambda\widehat{\rho}$ calculus (a version of Curien's $\lambda\rho$
calculus of explicit substitutions~\cite{Curien:TCS91} that is closed
under reduction~\cite{Biernacka-Danvy:TOCL07}) with a normal-order
reduction strategy and the Krivine machine~\cite{Swierstra:MSFP12}, which
included proving the refocusing step.
Also using Agda, Rozowski formalized the syntactic correspondence between
the $\lambda\widehat{\rho}$ calculus with an applicative-order reduction
strategy and the CEK machine~\cite{Rozowski:BSc}, which also included
proving the refocusing step.

\subsection{Applications of refocusing}
\label{sec:applications-of-refocusing}

The author's doctoral thesis~\cite[Section~3.2.1]{Danvy:DSc} lists
applications of refocusing until 2006:
to context-based CPS transformations,
to specifications of the $\lambda$ calculus with implicit and with explicit substitutions,
and
to specifications of the $\lambda$-calculus with explicit substitutions and computational effects,
with more examples listed at a summer school~\cite{Danvy:AFP08}.
Since then, refocusing has been used in a variety of settings including
normalization by evaluation~\cite{Danvy-Millikin-Munk:07,Munk:MS},
object-oriented programming~\cite{Danvy-Johannsen:JCSS10},
call by need~\cite{Danvy-al:TCS12,Danvy-Zerny:PPDP13},
combinatory graph reduction~\cite{Danvy-Zerny:TOCL13},
the SECD machine~\cite{Danvy-Millikin:LMCS08},
and
a block-structured imperative language~\cite{Zerny:PhD}.

{\c S}erb{\u a}nu{\c t}{\u a}, Ro{\c s}u, and Meseguer used refocusing for
the chemical abstract machine~\cite[Section~10]{Serbanuta-al:IaC09}.
Bach Poulsen and Mosses used refocusing to implement SOS and
MSOS~\cite{Bach-Poulsen-Mosses:LOPSTR13}.
As a component of the syntactic correspondence between calculi and
abstract machine, refocusing has been used
to derive type systems and implementations for coroutines~\cite{Anton-Thiemann:APLAS10},
to go from type checking via reduction to type checking via evaluation~\cite{Sergey:PhD},
to derive a classical call-by-need sequent calculus~\cite{Ariola-al:FLOPS12},
to derive interpretations of the gradually-typed lambda calculus~\cite{Garcia-Perez-al:PEPM14},
for full reduction~\cite{Garcia-Perez-Nogueira:SCP14},
and
to construct abstract machines for JavaScript~\cite{VanHorn-Might:2018}.

\section{Conclusion}
\label{sec:conclusion-and-perspectives}

Given a small-step semantics with a deterministic reduction strategy,
the point of refocusing is pragmatic: to deforest the reducts in
reduction sequences, \ie, to not construct them on the way from the site
of each redex to the site of the next redex:
{\let\labelstyle=\textstyle
\spreaddiagramrows{0.6cm}
\spreaddiagramcolumns{0.4cm}
 \newcommand{\atadless}{\hspace{-1mm}}
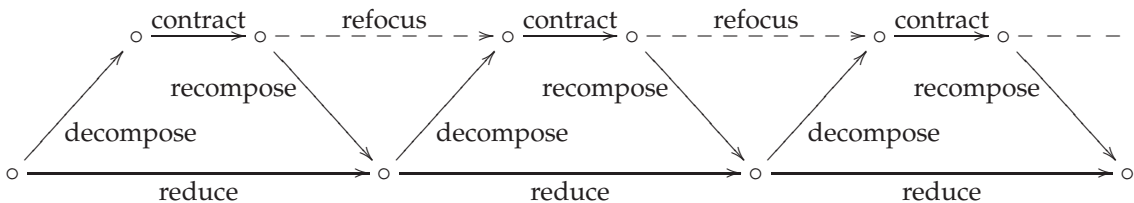
 $$
 \diagram
 &
 \circ
 \rto^{\text{{\small contract}}}
 &
 \circ
 \ar@{-->}[rr]^{\text{{\small refocus}}}
 \drto_>>>>>>>>>>>>>>>{\text{{\small recompose}}\atadless}
 &
 &
 \circ
 \rto^{\text{{\small contract}}}
 &
 \circ
 \ar@{-->}[rr]^{\text{{\small refocus}}}
 \drto_>>>>>>>>>>>>>>>{\text{{\small recompose}}\atadless}
 &
 &
 \circ
 \rto^{\text{{\small contract}}}
 &
 \circ
 \ar@{--}[r]
 \drto_>>>>>>>>>>>>>>>{\text{{\small recompose}}\atadless}
 &
 \\
 \circ
 \urto_>>>>>>>>>>>>{\atadless\text{{\small decompose}}}
 \ar@{->}[rrr]_{\text{{\small reduce}}}
 &
 &
 &
 \circ
 \urto_>>>>>>>>>>>>{\atadless\text{{\small decompose}}}
 \ar@{->}[rrr]_{\text{{\small reduce}}}
 &
 &
 &
 \circ
 \urto_>>>>>>>>>>>>{\atadless\text{{\small decompose}}}
 \ar@{->}[rrr]_{\text{{\small reduce}}}
 &
 &
 &
 \circ
 \enddiagram
 $$
}

In the early 2000's, Nielsen and the author identified the issue, which
was obvious, and observed that an optimal refocus function is simply the
decomposition function, which was less obvious.
Formally proving the correctness of refocusing was complicated because the
recomposing context is represented inside out and accumulated during
decomposition.
But contexts can be represented outside in too, and then structural
induction is enough for the proof to go through, witness
\sectionref{sec:the-simple-proof}.
And since the shortcut is achieved by a call to the decomposition
function, the result is as correct and as efficient as this decomposition
function.
In fact, experience has consistently shown that the resulting discovery
coincides with what we invent when we implement a reduction-free
normalization function,
\eg, in a definitional interpreter, for weak -- or for full -- normalization by
evaluation.

The proof presented here was a long time coming.
It came to be thanks to the availability of proof assistants.

\begin{flushright}
  \emph{There is a thin line between assumptions and conclusion.}
\end{flushright}

\paragraph{Acknowledgments:}
The author is grateful to Lasse R. Nielsen's for our initial joint work
about refocusing
and to Ma{\l}gorzata Biernacka for its continuation.
Thanks are also due to Julia Lawall for commenting on the first drafts of
this article in August 2022.

\bibliography{../../mybib.bib}
\bibliographystyle{plain}

\appendix

\section{Defunctionalization}
\label{app:defunctionalization}

In a given functional program, function spaces are only populated by
instances of the function abstractions that occur in this
program.
For example, consider
\inlinecoq{term -> term} in the
definitions of \inlinecoq{decompose_term_cc} and \inlinecoq{decompose_cc}
in
\sectionref{subsec:SOS-with-two-continuations-and-one-discontinuity:decomposition}:
\inputcoq{DECOMPOSE_TERM_KK}
\inputcoq{DECOMPOSE_KK}

\noindent
Only three function abstractions -- one in the initial call to
\inlinecoq{decompose_term_cc} and two in the tail-recursive calls to
\inlinecoq{decompose_term_cc} -- give rise to inhabitants of this
function space.
The idea of defunctionalization here is to represent this function space
as a ternary sum, with the first summand standing for the function
abstraction in the initial call and with the two others standing for the
function abstractions in the tail-recursive calls.
Furthermore each summand is first order and contains the denotations of
the free variables of the corresponding function abstraction:
\inputcoq{TERM_ARROW_TERM}

\noindent
Applying this representation of functions is carried out with a dispatch
function:
\inputcoq{DISPATCH_TERM_ARROW_TERM}

\noindent
This defunctionalization ripples out in the codomain of
\inlinecoq{decompose_term_cc}:
\inputcoq{VALUE_OR_DECOMPOSITION_CC_DEF}

\noindent
And here is the defunctionalized version of \inlinecoq{decompose_term_cc}
and \inlinecoq{decompose_cc}:
\inputcoq{DECOMPOSE_TERM_KK_DEF}
\inputcoq{DECOMPOSE_KK_DEF}
\inputcoq{REDUCE_KK_DEF}
\inputcoq{REDUCE_D_AND_REDUCE_CC_DEF_ARE_EQUIVALENT}

The type \inlinecoq{term_arrow_term} is isomorphic to the type of lists
of control frames:
\inputcoq{CONTROL_FRAME}
\inputcoq{CONTEXT}

\noindent
And so that is, mutatis mutandis, how we end with the defunctionalized
version of the decomposition function in
\sectionref{subsec:SOS-with-one-undelimited-continuation-and-one-context-and-one-discontinuity:decomposition}:
each function abstraction is replaced by a list construction, and each
function application is replaced by the application of a dispatch
function.
Since the function that is defunctionalized is a recomposing
continuation, its first-order type is the type of contexts and its
dispatch function is the recomposition function.

\clearpage

\small

\tableofcontents

\end{document}